\documentclass{elsart41}
\usepackage{graphics}
\usepackage{graphicx}
\usepackage{amssymb}
\usepackage{amsmath}
\usepackage{cite}
\usepackage[]{amsbsy}
\newcommand{\ff}[1]{{\boldsymbol #1}}
\begin{document}

\begin{frontmatter}



\title{Correlated Band Structure and the Ground-State Phase Diagram in
        High-T$_C$ Cuprates}
%

\author[GER]{Werner Hanke\corauthref{Name1}},
\ead{hanke@physik.uni-wuerzburg.de}
\author[AUS]{Markus Aichhorn},
\author[AUS]{Enrico Arrigoni},
\author[GER]{Michael Potthoff}

\address[GER]{Institute for Theoretical Physics, University of
        W\"urzburg, Am Hubland, 97074~W\"urzburg, Germany}  
\address[AUS]{Institute for Theoretical Physics and Computational
        Physics, Graz University of Technology, Petersgasse 16, 8010 Graz, Austria}

\corauth[Name1]{Corresponding author. Tel: +49 931 888-5714
fax: +49 931 888-5141}

\begin{abstract}
We review results obtained with a recently proposed variational
cluster approach (VCA) for the competition between d-wave
superconductivity (dSC) and antiferromagnetism (AF) in the high-T$_C$
cuprates. Comparing the single-particle spectra of a two-dimensional 
Hubbard model with quantum Monte-Carlo (QMC) and experimental data, 
we verify that the VCA correctly treats the low-energy excitations. 
The cluster calculations reproduce the overall ground-state phase diagram of the
high-temperature superconductors both for electron- and hole-doping. 
In particular, they include salient features such as the enhanced robustness 
of the AF state in case of electron doping. For electron- but also for hole-doping,
we clearly identify a tendency to phase separation into a mixed AF-dSC phase 
at low and a pure dSC-phase at high doping.

\end{abstract}

\end{frontmatter}


The central issue in the field of high-temperature superconductivity
(HTSC) is the connection of the microscopic interactions at the level of
electrons and ions, which are at high energy and temperature $T$, with
the ``emerging phenomena'' at $T=0$, i.e.\ competing and nearly degenerate 
orders -- antiferromagnetism (AF), d-wave superconductivity (dSC), heterogeneous 
phases, etc.
We will not go into a lengthy discussion of what interactions should be 
retained at the electron-ion level. But, when choosing the two-dimensional 
$(2D)$ one-band Hubbard model, i.e.\
\begin{equation}
        H = \sum_{i,j} t_{i,j} c^{\dagger}_{i} c_{j} +
        U \sum_{i} n_{i, \uparrow} n_{i, \downarrow} \: ,
\end{equation}
where $t_{i,j}$ denote hopping matrix elements, $n_{i,\uparrow}$ the density 
at site $i$ with spin ``$\uparrow$'' and $U$ the local Coulomb repulsion, 
one has introduced gross simplifications, leaving out other orbital (e.g.\ $p$)
degrees of freedom, long-range Coulomb interaction, electron-phonon coupling, 
etc \cite{Name1}.
Nevertheless, this model choice appears to be legitimate, last not
least in view of the amazing agreement achieved between numerical
simulations and experimental results for the normal-state properties
of the cuprates \cite{Name2, Name3}.

At low temperatures different orders appear, which are not 
separated by distinct energy scales but compete with each other.
What is required is a kind of ``magnifying lens'' which allows to resolve 
these competing orders. 
Ideally, one should employ a systematic renormalization-group approach to integrate 
out the irrelevant degrees of freedom and, thereby, to correctly bridge high to 
low energies and eventually to go to $T=0$.
For the strong-correlation case, realized in the HTSC, how to do this is, however, 
by no means obvious.
In this context, cluster techniques provide an alternative way to systematically 
approach the infinite-size (and, thereby, low-energy) limit.

Here, we review progress obtained with the variational cluster approach (VCA), 
which was proposed and used by Potthoff et.\ al.\ \cite{Name4, Name5}. 
This approach provides a rather general and controlled way to go to the 
infinite-sized lattice fermion system at low temperatures and at $T=0$, 
in particular.
The ground-state phase diagram of the $2D$ one-band Hubbard model was
calculated within VCA by S\'{e}n\'{e}chal et.\ al.\ \cite{Name7} and,
independently, by two of us \cite{Name6}. There are certain technical differences, 
which we discuss below,
but the ``upshot'' of the two works is as follows: 
For the cluster sizes used in the VCA, the $T=0$ phase diagram of the Hubbard 
model (1), with hopping terms up to third-nearest neighbors, correctly
reproduces salient features of the HTSC, such as the AF and dSC ground
states in doping ranges, which are qualitatively in agreement with electron- 
and hole-doped cuprates.

The VCA is based on the self-energy-functional approach (SFA) \cite{Pot03}. 
The SFA provides a variational scheme to use dynamical information from an 
exactly solvable ``reference system'' (for example an isolated cluster) to 
approximate the physics of a system in the thermodynamic limit. 
For a system with Hamiltonian $H=H_0(\ff t)+H_1(\ff U)$ and one-particle and 
interaction parameters $\ff t$ and $\ff U$, the grand potential is written as 
a functional of the self-energy $\ff \Sigma$ as
\begin{equation}
\Omega_{\ff t, \ff U} [\ff \Sigma] = F_{\ff U} [\ff \Sigma] + \mbox{Tr} \ln 
\left(\ff G_{0,\ff t}^{-1} - \ff \Sigma\right)^{-1} \: ,
\end{equation}
with the stationary property $\delta \Omega_{\ff t, \ff U} [\ff \Sigma_{\rm phys}] = 0$
for the physical self-energy. Here, $\ff G_{0, \ff t}=(\omega + \mu - \ff t)^{-1}$ is 
the free Green`s function at frequency $\omega$, and $\mu$ is the chemical potential. 
$F_{\ff U}[\ff \Sigma]$ is the Legendre transform of the Luttinger-Ward functional 
and determines the fully interacting Green`s function via 
$\ff G = -\delta F_{\ff U}[\ff \Sigma]/\delta \ff \Sigma$. 
It is important to note that $F_{\ff U}[\ff \Sigma]$ is a universal functional: 
The functional dependence is only determined by the interaction parameters $\ff U$
(for example, the Hubbard interaction in Eq.\ (1)).
Therefore, the functional $F_{\ff U}[\ff \Sigma]$ is the same as the functional for 
a problem which is ``simpler'' and solvable, i.e.\ for a Hamiltonian $H'= H_0(\ff t')+H_1(\ff U)$
with the same interaction part but a one-particle part that makes it {\em exactly}
solvable. 
The stationary solutions are obtained (and this is the approximation) {\em within} the 
subspace of self-energies $\ff \Sigma = \ff \Sigma(\ff t')$ of that simpler solvable 
problem that is spanned by varying $\ff t'$.
If one takes a single site and connects it to continuous (non-interacting) bath
degrees of freedom (another $H'$ choice), one recovers the dynamical mean-field theory
(DMFT) \cite{Name8} or, for a cluster of sites connected to a bath, a cluster
variant of DMFT \cite{Name9}.

In the VCA, considered in the following discourse, $H'$ is build up of disconnected clusters, 
which have their inter-cluster hopping terms removed. 
In our $T = 0$ approach, the isolated cluster is solved by exact diagonalization.
Its Hamiltonian $H'$ includes additional symmetry-breaking ``Weiss'' fields \cite{Name5} 
to allow for long-range order.
The VCA solution is finally obtained as a stationary point determined by 
$\partial \Omega_{\ff t,\ff U}[\Sigma (\ff t')] / \partial \ff t'=0$.
Is this a controlled route to a $(T = 0)$ infinite-size approach? To
answer this, consider a few ``tests'': 

\begin{figure}[t]
\begin{center}
   \includegraphics[width=\columnwidth]{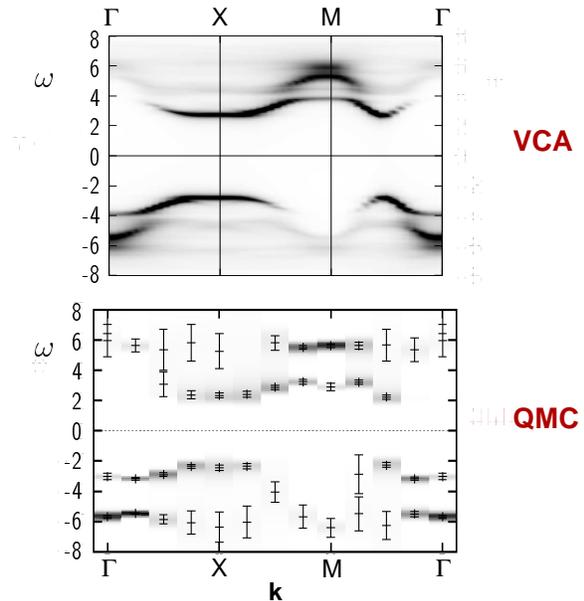}
    \caption{\label{1} 
Upper part: Density plot of the spectral function
for the $2D$ Hubbard model at half-filling, $T=0$ and $U=8t$ 
($t$: nearest-neighbor hopping) as obtained by the VCA \cite{Name5}. 
The lattice is covered by $\sqrt{10} \times \sqrt{10}$ clusters. 
Bottom: QMC (maximum entropy) result, taken from Ref. \cite{Name3}, 
for the same parameters but for a finite low temperature $T=0.1t$ 
and an isolated $8 \times 8$ cluster. Dark
(light) areas correspond to large (small) spectral weight.}
\end{center}
\end{figure}

(i) The VCA correctly reproduces long-range AF order in $2D$ and the absence 
of this order in $1D$ \cite{Name5}. This non-trivial test implies that the VCA 
goes well beyond ordinary mean-field theory. 

(ii) An advantage, compared to variational schemes based on wave functions \cite{Name10}, 
is that the VCA quite naturally gives the one-particle Green`s function $\ff G$. 
Fig.~\ref{1} compares the spectral function $A(\ff k, \omega) \propto \mbox{Im}
G(\ff k, \omega)$ of the VCA for the
$2D$ Hubbard model at $U=8t$, half-filling and $T=0$ with corresponding low-temperature 
QMC data \cite{Name3} for an isolated $8 \times 8$ cluster. 
One can clearly see that the VCA, with the lattice covered by $\sqrt{10} \times \sqrt{10}$ 
clusters, correctly reproduces coherent and incoherent ``bands'' (known from ARPES data 
\cite{Name11}). 
In particular, the non-trivial proliferation of AF spin correlations from cluster to cluster, 
which builds up the coherent quasi-particle ``band'' is obviously correctly embedded in the 
VCA \cite{Name5}. Similar calculations have been performed for the spectral function 
$A(\ff k,\omega)$ of the hole- and electron-doped Hubbard model \cite{Name7,Name6}. 
The characteristically different doping dependencies give rise to different Fermi-surface
evolutions upon doping.
Furthermore, the single-particle excitations provide insight into the characteristic 
differences in the ground-state phase diagram for hole- and electron-doping \cite{Name6}.

\begin{figure}[t]
\begin{center}
   \includegraphics[width=0.9\columnwidth]{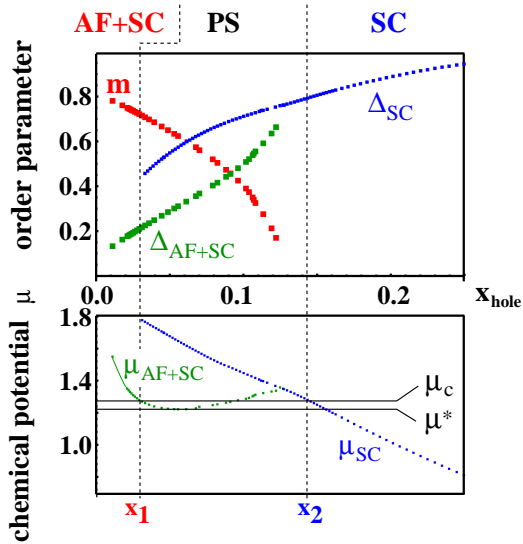}
\end{center}
\caption{\label{2} 
Antiferromagnetic and superconducting order parameters, $m$ and $\Delta$,
and chemical potential $\mu$ as functions of hole doping $x$. 
$\Delta$ and $\mu$ are plotted for the AF+SC 
(green, $\Delta_{\rm AF+SC}$, $\mu_{\rm AF+SC}$) as well
as for the pure SC homogeneous solutions (blue, $\Delta_{\rm SC}$, $\mu_{\rm SC}$). 
Note that $\Delta$ is scaled by a factor 10 for convenience. 
For $x<x_1$ the system exhibits a coexistence of AF and dSC order.
Phase separation (PS) occurs between the doping levels $x_1$ and $x_2$.
For $x>x_2$ pure dSC is realized.
In the phase separation region $x_1 < x < x_2$, the homogeneous solutions 
become unstable, and the system prefers to separate into a mixture of two 
densities corresponding to $x_1$ and $x_2$. 
The chemical potential $\mu_c$ is determined by the Maxwell construction 
shown in the figure. 
At $\mu^\ast$ the slope of the AF+SC solution changes sign.
}
\end{figure}

\begin{figure}[b]
\begin{center}
        \includegraphics[width=0.98\columnwidth]{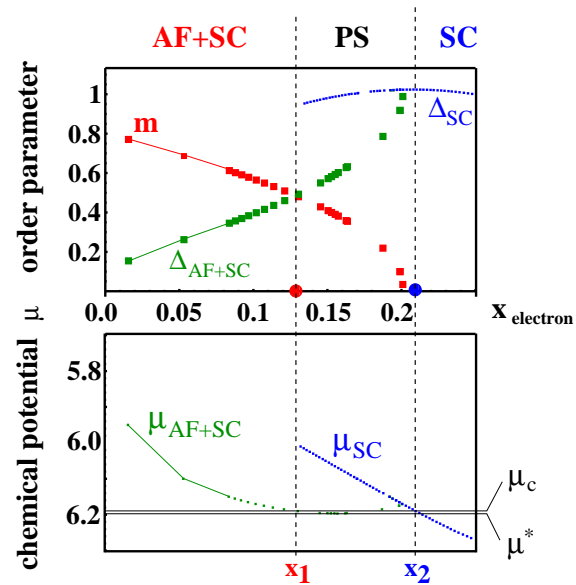}
\end{center}
\caption{\label{3} 
The same as Fig.~\ref{2} but for electron doping.
Note the enhanced robustness of the AF state and the 
strongly reduced scale 
$\Delta \mu \equiv (\mu^\ast - \mu_{c})$ as compared to hole doping.
}
\end{figure}
	
(iii) 
To test the stability of the homogeneous phases with respect to phase separation (PS), we
consider a reference system $H'$ of isolated $2\times 2$ clusters where, in addition to 
the two symmetry-breaking terms (Weiss fields) $H'_{\rm AF}$ and $H'_{\rm SC}$, a term
$H'_{local}$ is optimized within the variational procedure \cite{Name6}.
$H'_{local} = \varepsilon \sum_{i\sigma} n_{i\sigma}$ describes a shift $\varepsilon$ 
of the chemical potential in the cluster with respect to the physical chemical potential $\mu$.
The use of the additional variational parameter $\varepsilon$ is required in order to
have a consistent treatment of the particle density.
The optimization of $\varepsilon$ has to be done simultaneously with the optimization 
of the parameters $h_{\rm AF}$ and $h_{\rm SC}$, namely the staggered magnetic field 
in the term $H'_{\rm AF}$ and the nearest-neighbor d-wave pairing field $h_{\rm SC}$
in the term $H'_{\rm SC}$.
Notice that the Weiss fields $h_{\rm AF}$ and $h_{\rm SC}$ are different from the 
corresponding order parameters $m$ and $\Delta$ plotted in Figs.~\ref{2} and \ref{3}.
Quite generally, however, a nonvanishing stationary value for the Weiss fields produces
a nonvanishing order parameter, although the latter can be much smaller.

The phase diagram for the Hubbard model with $U=8t$ and next-nearest-neighbor hopping
$t_{n.n.n}=-0.3 t$, obtained with our calculation, is plotted in Fig.~\ref{2} for the 
hole-doped and in Fig.~\ref{3} for the electron-doped case.
In the upper part of each figure, we display the AF $(m)$ and dSC
$(\Delta)$ order parameters as a function of doping $x$.
In the lower part of the figures, the chemical potential $\mu$ is
plotted as a function of $x$. 

Let us discuss hole doping first (see Fig.~\ref{2}).
For dopings $x$ below a critical value $x_1$ we find a homogeneous symmetry-broken 
state in which both, the AF as well as the dSC order parameter $m$ and $\Delta$ are 
non-zero.
This corresponds to a phase AF+SC where AF and dSC order microscopically and coherently 
coexist.
A homogeneous state with pure dSC ($m=0$ and $\Delta >0$) is obtained for dopings $x>x_2$.

Fig.~\ref{2} also shows $\Delta$ and $\mu$ for the homogeneous AF+SC and SC phases in 
the range $x_1 < x < x_2$. 
Here, however, these phases are thermodynamically unstable. 
For dopings $x$ with $x_1 < x < x_2$, macroscopic phase separation between the two 
phases occurs.
In practice, doping-dependencies are calculated by varying the chemical potential $\mu$. 
Following up the grand potentials for the two homogeneous phases as functions of $\mu$,
i.e.\ $\Omega_{\rm AF+SC}(\mu)$ and $\Omega_{\rm SC}(\mu)$, it is found (see 
Ref.~\cite{Name6} for details) that there is a crossing at a critical chemical potential
$\mu=\mu_c$ (at this point the AF order parameter $m$ is still nonzero). 
Thus, the transition is first order as a function of $\mu$. 
At the transition point $\mu_{c}$, the dopings corresponding to the AF+SC and to the SC
phase, $x_{\rm AF+SC}$ and $x_{\rm SC}$, are different.
Consequently, there is a jump $\Delta x \equiv x_{\rm AF+SC} - x_{\rm SC}$ at $\mu_{c}$, 
indicating phase separation between a weakly doped AF+SC and a higher doped SC phase. 

Due to the inclusion of $H'_{\rm local}$, $\mu_{c}$ can {\em equivalently} be obtained
by a Maxwell construction. This is shown in 
Fig.~\ref{2} where, in the lower panel, $\mu$ is plotted as a function of $x$.
Here, phase separation is signaled by the fact that the $\mu(x)$ is not a monotonous 
function. 
The Maxwell construction shown in the figure then identifies the two dopings $x_1$ and 
$x_2$ into which the system tends to phase separate, as well as the chemical potential 
$\mu_c$ in the phase-separated state.
In Fig.~\ref{2}, $\mu^\ast$ is the point where the slope of $\mu(x)$ changes
sign. 
For $\mu < \mu^\ast$ the AF+SC solution ceases to exist

Let us now discuss the electron-doped case.
While the phase diagrams in Figs.~\ref{2} and \ref{3} are {\em qualitatively} similar, 
the phase in which long-range AF order is realized is spreading to significantly 
larger doping values in the electron-doped case, in overall agreement with the experimental
situation.
Another important difference concerns the energy scale for phase separation, 
i.e.\ $\Delta \mu \equiv (\mu^\ast - \mu_{c})$.
As one can see from the comparison between Figs.~2 and 3, $\Delta\mu$ is an order of 
magnitude larger in the hole-doped case. 
In Ref.~\cite{Name6} it is argued that this can explain the different
pseudogap and SC transition scales in hole- and electron-doped materials.
This may give support to theories \cite{Name12} which are based on the idea that
fluctuations of the competing phases, or of the related order parameters, are responsible 
for the pseudogap phenomenon.

In order to resolve the relevant small energy scale, it is necessary to evaluate $\Omega$ as 
well as its stationary points with high accuracy. Furthermore, the inclusion of the chemical 
potential shift term $H'_{local}$ considerably complicates the variational optimization.
For the rather small clusters of size $2 \times 2$ used here, the reference system can
be treated by full diagonalization and the frequency integrals, which are implicit in 
Eq.\ (2) \cite{Pot03}, can be carried out by means of a sum over the negative poles of 
the Green's functions. 
S\'{e}n\'{e}chal et.\ al.\ \cite{Name7} have considered clusters up to 10 sites and
report similar results for $x \approx 0$ but, without the inclusion of $H'_{local}$, 
one cannot reliably test the stability of the homogeneous phases against phase separation. 
Note that the transition from the AF+SC to the SC phase may appear as continuous as 
a function of $x$ if phase separation is not taken into account.

In conclusion, there has been substantial recent progress in relating the ``high-energy'' 
physics of the Hubbard model and its variants to the low-energy physics of the competing 
phases AF, dSC, charge inhomogenities, etc. 
This progress is due to the development of quantum-cluster theories, such as the VCA discussed here but 
also due to cluster extensions of the DMFT, such as the dynamical cluster approximation 
(DCA) \cite{Name13, Name14}.
Within these cluster approaches, characteristic 
difficulties have been encountered: The latest impressive work using
the DCA by Maier et.\ al.\ performed a systematic cluster-size study of
dSC in the $2D$ Hubbard model \cite{Name13}. In clusters large enough
(up to 26 sites) converged results point to a finite-$T$ instability to
dSC. Because of the QMC minus-sign problem, however, results were
limited to $U=4t$, where the typical energy separation in $U$ and the
magnetic energy scale of the HTSC is not yet achieved. On the other hand,
the VCA studies reviewed here, are clearly not yet converged with respect to 
the cluster size, as one can read off from Fig. 1 in Ref \cite{Name7}. 
Cluster convergence is, at least in principle, also possible within the $VCA$.
With increasing cluster size longer-ranged correlations are included exactly.
This, however, necessarily implies to use stochastic (QMC) methods as solvers
for the cluster reference system.
\\

One of us (WH) would like to acknowledge the hospitality of the Kavli
Institute for Theoretical Physics in Santa Barbara, where part of this
work was finished (supported by Nat. Sc. Found. Grant No. PHY99-0794).
We would like to thank D. J. Scalapino for many useful discussions.
This work was also supported by the DFG Forschergruppe 538, by the KONWIHR
supercomputing network in Bavaria, and by
the Doctoral Scholarship Program of the Austrian Academy of Sciences.

\end{document}